\documentclass[12pt]{article}
\usepackage{epsfig,cite,a4wide}
\newcommand{\beq}{\begin{equation}}
\newcommand{\eeq}[1]{\label{#1}\end{equation}}
\newcommand{\ra}{\rightarrow}

\newcommand{\be}{\begin{equation}}
\newcommand{\ee}{\end{equation}}
\newcommand{\ba}{\begin{eqnarray}}
\newcommand{\ea}{\end{eqnarray}}

\begin{document}

\title{$\Theta^+$ Hypernuclei}

\author{D. Cabrera, Q.B. Li, V.K. Magas, E. Oset and M.J. Vicente Vacas\\
%}\affiliation{
{\small
 Departamento de F\'{\i}sica Te\'orica and IFIC, Centro Mixto }\\
{\small  Universidad de Valencia - CSIC, Institutos de Investigaci\'on de Paterna, }\\ 
{\small  Apdo. correos 22085, 46071, Valencia, Spain}
}

\date{\today}

\maketitle

\begin{abstract}
%{
%Abstract: 
We present results for the selfenergy of the $\Theta^+$ pentaquark in nuclei
associated with two sources: the $KN$ decay of the $\Theta^+$ and the two meson
baryon
decay channels of the $\Theta^+$ partners in an antidecuplet of baryons. The
first source is shown to produce a small potential, unable to bind the 
$\Theta^+$ in nuclei, while the second source gives rise to a large attractive
potential.  At the same time we show that the width of the $\Theta^+$ in nuclei
is small, such that, in light and medium nuclei, many bound 
$\Theta^+$ states would appear with a separation between levels appreciably 
larger than the width of the states, thus creating an ideal scenario for
pentaquark spectroscopy in nuclei.

%}
\end{abstract}

\section{Introduction}
\label{int}

The physics of hypernuclei, $\Lambda$, $\Sigma$, $\Xi$ is one of the active
branches of nuclear physics with steady progress at the experimental and
theoretical levels 
\cite{Dover:sv,Oset:1989ey,Oset:1998xz,Alberico:2001jb,Gal:si,Hashimoto:1999ks,Itonaga:2002gj}.
It has brought information on the $\Lambda N$ interaction, the $\Lambda N \ra
NN$ weak transition, interesting examples of drastic Pauli blocking effects in
the  $\Lambda \ra \pi N$ decay in nuclei, one of the cleanest examples of the
accuracy of the mean field approximation in the case of $\Lambda$ hypernuclei,
a striking example of medium effects with increases of a factor fifty or more
in the mesonic $\Lambda$ decay width due to the interaction  of the pion with
the nucleus and other topics. So far only hypernuclei with strangeness $-1$ or
$-2$ have been formed.  The discovery of an exotic baryon with positive
strangeness, $\Theta^+$ \cite{Nakano:2003qx} (see also Ref. \cite{hyodo} for a
list of experimental and theoretical related works), opens new possibilities of
forming exotic $\Theta^+$ hypernuclei which, like in the case of negative
strangeness hypernuclei, can provide information  unreachable or complementary
to that obtained in elementary reactions.

Suggestions that $\Theta^+$ could be bound in nuclei have already been made. In
Ref. \cite{Miller:2004rj} a schematic model for quark-pair interaction with
nucleons was used to describe the $\Theta^+$, which suggested that $\Theta^+$
hypernuclei, stable against strong decay, may exist.  In Ref. \cite{Kim:2004fk}
the $\Theta^+$ selfenergy in the nuclei is calculated, based on the $\Theta^+
\to K N$ decay mode, Pauli blocking and a mass modification of the nucleon in
the nuclear matter. The resulting selfenergy is too weak to bind the $\Theta^+$
in nuclei.

In the present work we redo the calculations of Ref. \cite{Kim:2004fk}
modifying the assumption of a strong shift of the nucleon mass and
renormalizing the kaon cloud in the nucleus. The results are qualitatively
similar to those of Ref. \cite{Kim:2004fk}
and a small potential is obtained from this source.  As a novelty, we
also evaluate the imaginary part of the potential and show that the $\Theta^+$
width becomes smaller for possible nuclear bound states and would be narrow
enough to allow distinct peaks to be seen experimentally, provided some large
attraction is obtained from other source. This is the other issue we deal with
in this work. Indeed, we show that the in-medium renormalization of the pion in
the two meson cloud of the $\Theta^+$ leads to a sizable attraction, enough to
produce a large number of  bound and narrow $\Theta^+$ states in nuclei.

The coupling of the $\Theta^+$ to two mesons and a nucleon is studied in Ref.
\cite{madrid} where, with the assumption that the $N^*(1710)$ resonance has a
large component in the same antidecuplet as the $\Theta^+$,
two terms of a $SU(3)$ symmetric
Lagrangian are constructed to account for the $N^*\ra N$ ($\pi\pi$, $p-$wave,
$I=1$) and  $N^*\ra  N$ ($\pi\pi$, $s-$wave, $I=0$) partial decay widths of the
$N^*(1710)$. With this  Lagrangian an attractive selfenergy is obtained for
all the  members of the antidecuplet coming from the two meson cloud.

In this work we study the nuclear
medium effects on the $\Theta^+$ selfenergy diagrams derived from this
$\Theta^+ \ra K\pi N$ Lagrangian. This is accomplished by modifying the pion,
kaon and nucleon propagators in the nuclear medium.  We find quite a large
attractive potential of the $\Theta^+$ which leads to bound states even for
light nuclei. We also investigate a new source of $\Theta^+$ decay width,
namely $\Theta^+\ra N K ph$, where the $ph$ (particle-hole) comes from the
absorption of a virtual pion, and we find it to be rather small. Altogether,
the total in-medium $\Theta^+$ width is much smaller than the separation of the
deeper $\Theta^+$ energy levels that we obtain for most nuclei, which could
open the grounds for $\Theta^+$ spectroscopy in nuclei.

\section{The $\Theta^+$ selfenergy from $KN$ decay channel} 
\label{Sigma_korean}

\begin{figure}
\begin{center}
    \includegraphics[height=5.0cm]{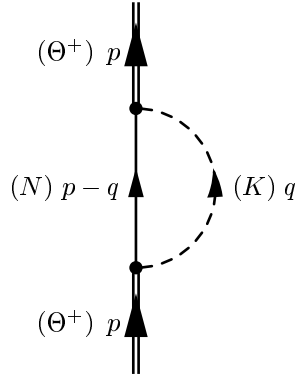}
\caption{$\Theta^+$  selfenergy diagram related to the $K N$ decay channel.}
\label{2bodyself}
\end{center}
\end{figure}

We begin by evaluating the selfenergy of the $\Theta^+$ related to the $KN$
decay channel in the medium. The $\Theta^+$  selfenergy diagram is depicted in Fig.
\ref{2bodyself}.

We assume first $I=0$ and $J^P=\frac{1}{2}^-$ for the $\Theta^+$. This implies
an $L=0$ coupling to $KN$. The
$KN$ state in $I=0$ is
\beq
|KN,I=0>=\frac{1}{\sqrt{2}}(|K^+n>-|K^0p>)\,. 
\eeq{eq1}
The $\Theta^+KN$ couplings in this case are
\beq
-i\,t_{\Theta^+K^+n}=-ig_{K^+n};\quad -i\,t_{\Theta^+K^0p}=ig_{K^+n}\,.
\eeq{eq2} 

For the $I=0$, $L=1$, $J^P=\frac{1}{2}^+$ case, the quantum numbers of the
antidecuplet suggested in Ref. \cite{Diakonov:1997mm}, we would have  
\beq
-i\,t_{\Theta^+K^+n}=-\bar{g}_{K^+n}\vec{\sigma}\vec{q};\quad
-i\,t_{\Theta^+K^0p}=\bar{g}_{K^+n}\vec{\sigma}\vec{q}\,, 
\eeq{eq3}
with $q$ the outgoing kaon momentum. This amplitude is a
nonrelativistic reduction of the relativistic vertices used in Ref. \cite{Kim:2004fk}. We
have also done the complete relativistic calculations and the differences are
negligible.

In the case of $L=1$ we could also have $J^P=\frac{3}{2}^+$ and the couplings
are written in terms of the corresponding spin transition operators, but it is
easy to see, following the steps of \cite{Hyodo:2003jw}, that the results for
the selfenergy would be the same as in the $J^P=\frac{1}{2}^+$ case. Similarly,
we could also assume $I=1$, which would only change the relative sign of the
$KN$ components in Eq. (\ref{eq1}), which appear squared in the selfenergy.
Thus, the $\Theta^+$ selfenergy does not change by assuming $I=0$ or $1$. We
have only two independent cases, $L=0$ and $L=1$, which we evaluate below.

For the $L=0$ case the free $\Theta^+$ selfenergy from the diagram in Fig.
\ref{2bodyself} is given by
\beq
-i\Sigma_{KN}(p)=2\int\frac{d^4q}{(2\pi)^4} (-ig_{K^+n})^2 
\frac{M}{E_N(\vec{p}-\vec{q})}\,
\frac{i}{p^0-q^0-E_N(\vec{p}-\vec{q})+i\epsilon}\,
\frac{i}{q^2-m_K^2+i\epsilon}\,,
\eeq{eq4} 
where $M$ is the nucleon mass, $E_N(k)=\sqrt{M^2+\vec{k}\,^2}$, and the factor
2 accounts for the  $K^+n$ and $K^0p$ channels, which leads to a $\Theta^+$
decay width
\beq
\Gamma=-2\,\textrm{Im}\, \Sigma_{KN} 
= \frac{g_{K^+n}^2}{\pi}\frac{q_{on}}{M_{\Theta^+}}\,,
\eeq{eq5}
where $q_{on}$ is the momentum of the kaon in the $\Theta^+\ra KN$ decay. The
result for $L=1$ is obtained by the substitution 
$$
g_{K^+n}^2\ra \bar{g}_{K^+n}^2 \vec{q}\,^2 \, ,
$$
and hence
\beq
\Gamma=\frac{\bar{g}_{K^+n}^2}{\pi}\frac{q_{on}^3}{M_{\Theta^+}}\,.
\eeq{eq6} 
  
\begin{figure}
\begin{center}
\includegraphics[width=0.7\textwidth]{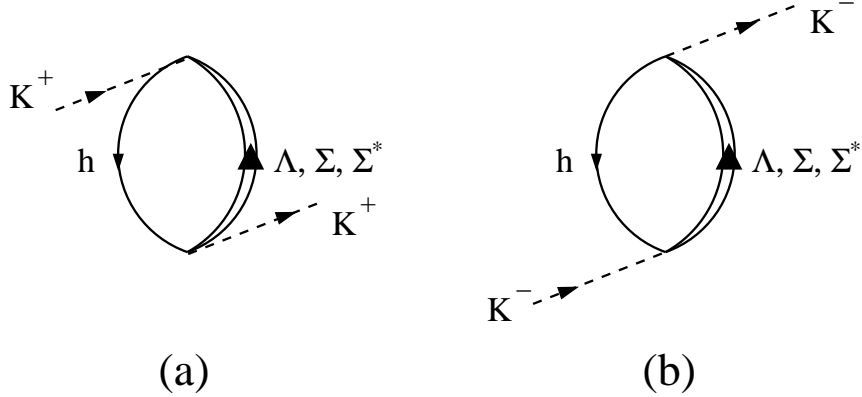}
\caption{\label{PwaveKself} In-medium kaon $p-$wave selfenergy diagrams: (a) $K$ crossed
term; (b) $\bar{K}$ direct term.}
\end{center}
\end{figure}

We proceed now to evaluate the $\Theta^+$  selfenergy in an infinite nuclear
medium with density $\rho$. First, the nucleon propagator changes in the
following way,
\beq
\frac{1}{p^0-q^0-E_N(\vec{p}-\vec{q})+i\epsilon}\ra 
\frac{1-n(\vec{p}-\vec{q})}{p^0-q^0-E_N(\vec{p}-\vec{q})+i\epsilon}\,+\,
 \frac{n(\vec{p}-\vec{q})}{p^0-q^0-E_N(\vec{p}-\vec{q})-i\epsilon}\,,
\eeq{eq7}
where $n(\vec{k})$ is the occupation number of the uncorrelated Fermi sea. On
the other hand, the vacuum kaon propagator is replaced by the in-medium one,
\beq
\frac{1}{q^2-m_K^2+i\epsilon}\ra\frac{1}{q^2-m_K^2-\Pi_K(q,\rho)}\,,
\eeq{eq8}
where $\Pi_K(q^0,|\vec{q}|,\rho)$ is the kaon selfenergy which accounts for
$s-$wave and $p-$wave $KN$ interaction. The $s-$wave part of the self energy is
well approximated by \cite{Kaiser:1996js,Oset:2000eg} 
\beq
\Pi_{K}^{(s)}(\rho)=0.13\, m_K^2 \rho/\rho_0\ \ [\textrm{MeV}^2]\,,
\eeq{eq9} 
where $\rho_0$ is the normal nuclear density. The $p-$wave part is taken such
that
\beq
\Pi^{(p)}_{K^+}(q^0,|\vec{q}|,\rho)=\Pi^{(p)}_{K^-}(-q^0,|\vec{q}|,\rho) \,,
\eeq{eq10} 
and for $\Pi_{K^-}^{(p)}$ we take the model of Refs. \cite{Oset:2000eg,Cabrera:2002hc}
which accounts for $\Lambda h$, $\Sigma h$ and $\Sigma^*(1385) h$ excitations,
see Fig. \ref{PwaveKself}.
Since the $p-$wave selfenergy of the $K^+$ involves crossed terms of the $Y h$
excitation, this part is small and there is practically no $q^0$ dependence in
the $K^+$ selfenergy, which makes the quasiparticle approximation accurate.
Hence, the pole structure of the free $K^+$ propagator, with poles in
$q^0=\pm\, \omega(q)\mp i\epsilon$, is substituted by a similar one with the
poles shifted to $\widetilde{\omega}(q)$, such that
\begin{equation}
\label{Kaondispersion}
\widetilde{\omega}(q)^2 - \vec{q}\,^2 - m_K^2 -
\Pi_K(\widetilde{\omega}(q),|\vec{q}|,\rho) = 0 \,.
\end{equation}
This equation is solved selfconsistently and the result is very close to
\begin{equation}
\label{KaondispersionnoPwave}
\widetilde{\omega}(q) \simeq \sqrt{m_K^2 (1+0.13 \rho/\rho_0)+\vec{q}\,^2}
\,.
\end{equation}
In view of this, the $q^0$ integration is  performed in the modified Eq.
(\ref{eq4}) leading to
\begin{eqnarray}
\label{SigmaAll}
& &\Sigma_{KN} (p^0,\vec{p};\rho) = 
\nonumber \\
&=& \frac{M_{\Theta} \Gamma}{M q_{on}} 
\frac{1}{(2\pi)^2} \int d^3 q \frac{M}{E_N(\vec{p}-\vec{q})} {\cal F}_L (q)
\frac{1}{2\widetilde{\omega}(q)} \,
\frac{1}{p^0-\widetilde{\omega}(q)-E_N(\vec{p}-\vec{q})-V_N+i\epsilon}
\nonumber \\
&-& \frac{M_{\Theta} \Gamma}{M q_{on}} \frac{1}{(2\pi)^2}
\int d^3 q \frac{M}{E_N(\vec{p}-\vec{q})} {\cal F}_L (q)
\frac{n(\vec{p}-\vec{q})}
{[p^0-E_N(\vec{p}-\vec{q})-V_N]^2-\vec{q}\,^2-m_K^2-\Pi_K(q,\rho)}
\end{eqnarray}
with $q^0=p^0-E_N(\vec{p}-\vec{q})-V_N$, $V_N=-\frac{k_F^2}{2M}$, and ${\cal
F}_0=1$, ${\cal F}_1=\frac{\vec{q}\,^2}{q_{on}^2}$.  In Eq. (\ref{SigmaAll}) we
have also taken into account the nucleon binding and, consistently with the
posterior use of the results within the local density approximation ($\rho \to
\rho(r)$ in the nucleus), we have taken the Thomas-Fermi potential for the
nucleons, $V_N=-k_F(r)^2/2M$ with $k_F(r)=(3\pi^2\rho(r)/2)^{1/3}$. For the
calculations we have taken an average value of the momentum of the $\Theta^+$
in eventual bound states of $p=200$ MeV, similar to that of bound nucleons in
nuclei.

The consideration of the $\Theta^+$ momentum is important for the imaginary
part of $\Sigma_{KN}$, and hence for the width
of the $\Theta^+$ in the nucleus. This can be easily understood.  Consider  a
$\Theta^+$ at rest with $1540$ MeV of energy decaying into $K^+ n$. The
momentum of the neutron is $269.6$ MeV/c. Taking $\rho=\rho_0$, $k_F=269$ MeV/c
and hence the decay would be allowed.  However, if the $\Theta^+$ has a small
initial momentum, it is clear  that when boosting the neutron momentum from the
$\Theta^+$ rest frame to the $\Theta^+$ moving frame, approximately  half of
the events would lead to $p_n > k_F$ while the other half would lead to $p_n <
k_F$, hence reducing the $\Theta^+$ width to one half its free value.

\begin{figure}
\begin{center}
    \includegraphics[width=0.75\textwidth]{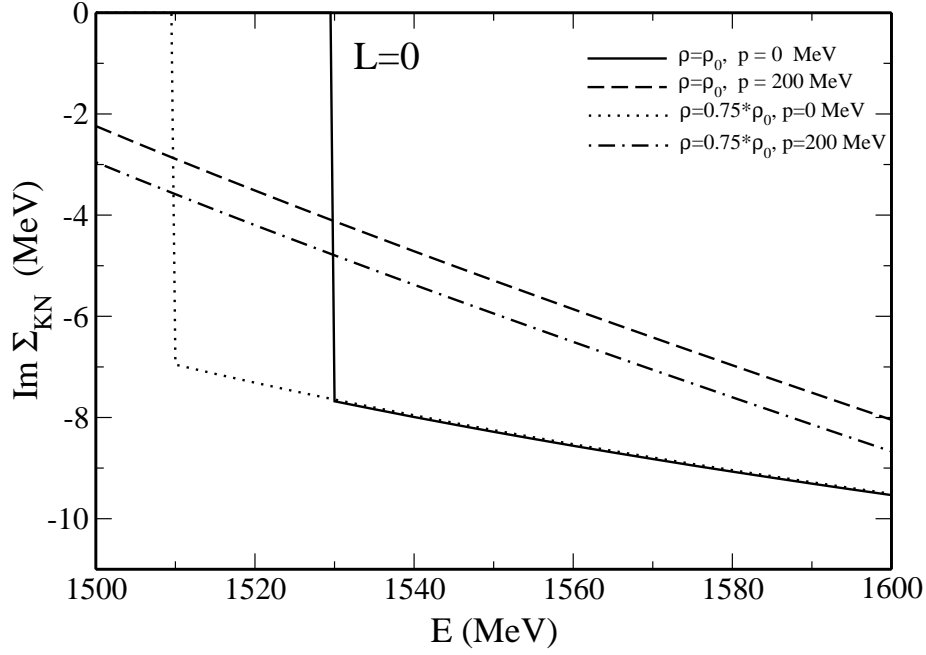}
\caption{Imaginary part of the $\Theta^+$ selfenergy associated to the $KN$ decay
channel for $L=0$.}
\label{ImL0}
\end{center}
\end{figure}

\begin{figure}
\begin{center}
    \includegraphics[width=0.75\textwidth]{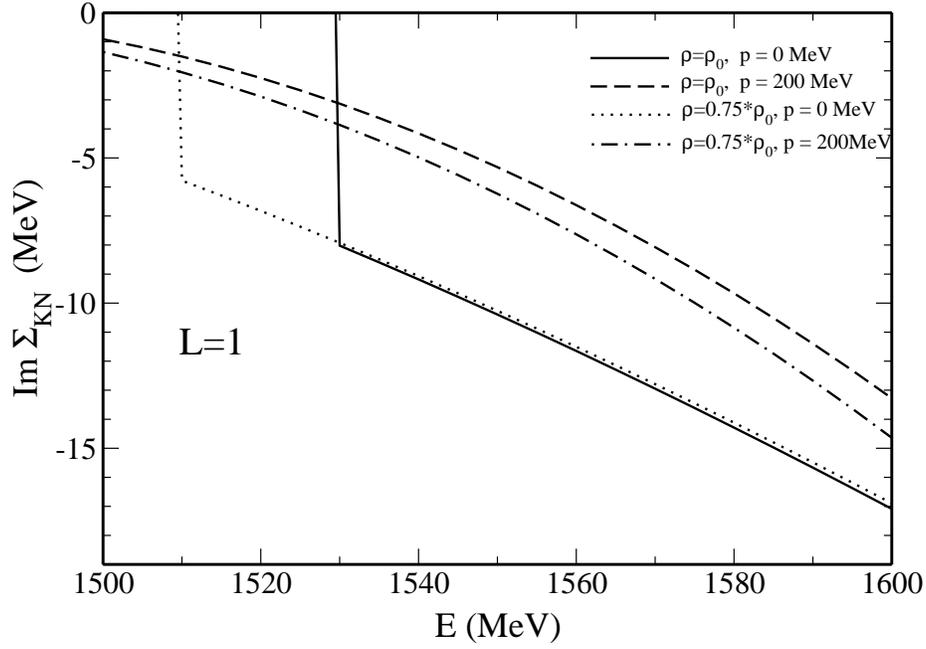}
\caption{Imaginary part of the $\Theta^+$ selfenergy associated to the $KN$ decay
channel for  $L=1$.}
\label{ImL1}
\end{center}
\end{figure}

This result is roughly valid for any momentum of the $\Theta^+$ smaller than
$k_F$. We should expect a reduction by a factor around two of the $\Theta^+$
width only from this source. On the other hand, if the $\Theta^+$ energy is
smaller because it is in a bound state, then the width is further reduced. We
can see this in Figs. \ref{ImL0}, \ref{ImL1} where we assume that $\Gamma=15$
MeV. This is the upper limit in most experiments coming basically from the
experimental resolution. Studies based on $K^+ N$ scattering suggest that the
width should be smaller than $5$ MeV
\cite{Nussinov:2003ex,Arndt:2003xz,Haidenbauer:2003rw} or even of
the order of $1$ MeV \cite{Cahn:2003wq,Gibbs:2004ji,Sibirtsev:2004bg}. What we see in these
figures is that even if $\Gamma=15$ MeV in free space, inside the nucleus,
particularly for the case of $L=1$ which corresponds to $J^P=\frac{1}{2}^+$ for
the $\Theta^+$, the width is small; and for $20$ MeV of $\Theta^+$ binding the
width would go down from $15$ MeV to less than $6$ MeV. This width could be
reasonably smaller than the separation between different bound levels.

\begin{figure}
\begin{center}
    \includegraphics[width=0.75\textwidth]{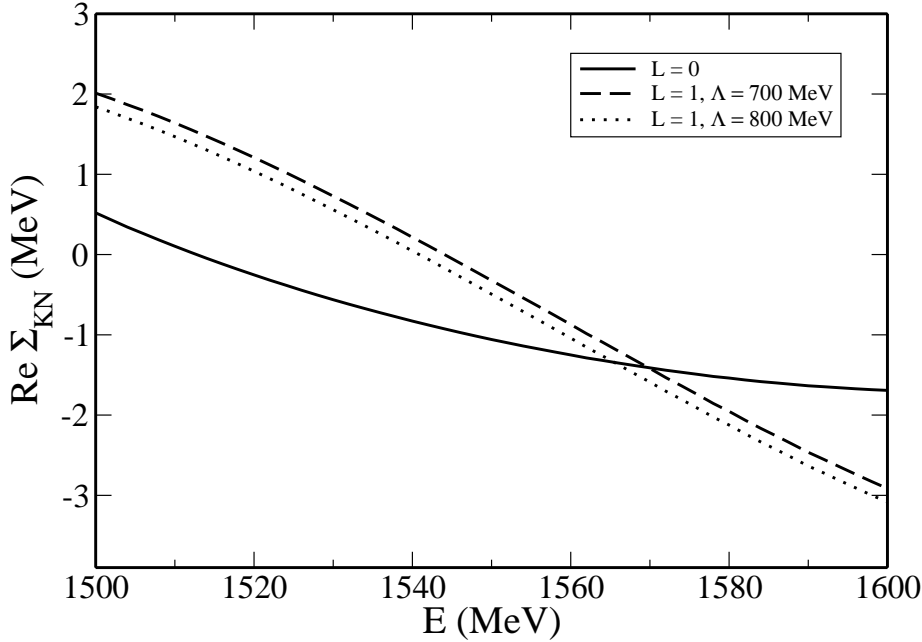}
\caption{Real part of the $\Theta^+$ selfenergy associated to the $KN$ decay
channel at $\rho=\rho_0$. A momentum of the $\Theta^+$ of 200 MeV is taken.}
\label{Realpart}
\end{center}
\end{figure}

Next, we consider the real part of the $\Theta^+$ selfenergy, shown in Fig.
\ref{Realpart}, after subtracting the vacuum selfenergy from Eq. (\ref{eq4}), for
a typical finite momentum of 200 MeV.  The subtraction is convergent for the
$L=0$ case, whereas for $L=1$ we use a cut-off in the momentum of the particles
in the loop.  As shown in the figure, the cut-off dependence is small. We find,
in qualitative agreement with Ref. \cite{Kim:2004fk}, that the $\Theta^+$
potential in the medium is very small. Note however that the results are not
directly comparable since we present the real part of the selfenergy, instead
of the in-medium $\Theta^+$ mass change presented in Ref. \cite{Kim:2004fk}.
According to Eq. (\ref{SigmaAll}) the selfenergy scales like $\Gamma$, which we
have taken as 15 MeV for the results shown for both the real and imaginary
parts of the selfenergy and if, as suggested in
\cite{Cahn:2003wq,Gibbs:2004ji,Sibirtsev:2004bg}, the width is of the order of 1 MeV, the
in-medium selfenergy associated to the $KN$ decay channel would be negligible. 
In any case, up to $\rho=\rho_0$ the real part of the selfenergy is not enough
to bind $\Theta^+$ in nuclei.

In the next section we investigate another source of attraction which leads to
larger attractive potentials.

\section{The $\Theta^+$ selfenergy tied to the two-meson cloud}
\label{3body}

\begin{figure}
\begin{center}
\includegraphics[width=0.7\textwidth]{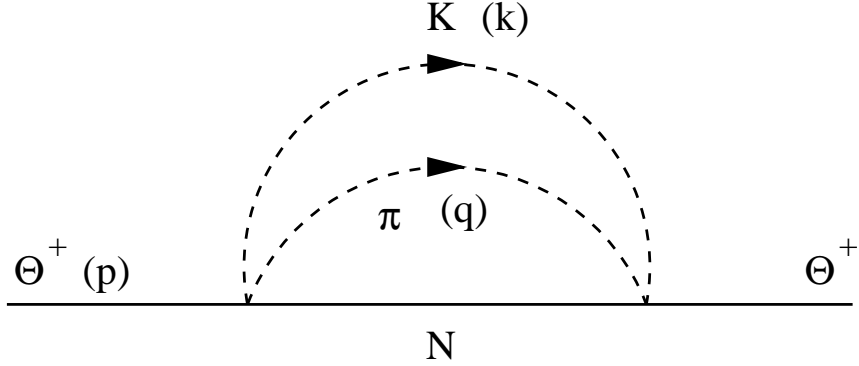}
\caption{\label{Threebody} Two-meson $\Theta^+$ selfenergy diagram.}
\end{center}
\end{figure}

In this section
we will study contributions to the $\Theta^+$ selfenergy from diagrams in which
the $\Theta^+$ couples to a nucleon and two mesons, like the one in Fig.
\ref{Threebody}.
There is no direct information on these couplings since the $\Theta^+$ mass is
below the two-meson decay threshold.

From now on we will do several assumptions. The validity of our results depends
on them. First, the $\Theta^+$ is assumed to have $J^P=1/2^+$ associated to an
$SU(3)$ antidecuplet, as in Ref. \cite{Diakonov:1997mm}. In addition, the
$N^*(1710)$ is supposed to couple largely to this antidecuplet.

From the data on $N^*(1710)$ decays we can determine the couplings to the
two-meson channels, and using $SU(3)$ symmetry obtain the corresponding couplings for
the $\Theta^+$.

In Ref.~\cite{madrid} two $SU(3)$ symmetric Lagrangian terms, with minimal
number of derivatives in the meson fields,
are proposed in order to account for the 
$N^*(1710)$ decay into $N (\pi\pi, p-\textrm{wave},I=1)$ and $N (\pi\pi,
s-\textrm{wave},I=0)$. The first term is
\begin{equation}
\label{L1}
{\cal L}_1 = i g_{\bar{10}} \epsilon^{ilm} \bar{T}_{ijk} \gamma^{\mu} B^j_l
(V_{\mu})^k_m \, ,
\end{equation}
with $V_{\mu}$ the vector current which for two mesons is written as
\begin{equation}
\label{veccurr}
V_{\mu} = \frac{1}{4 f^2} (\phi \partial_{\mu} \phi - \partial_{\mu} \phi \phi)
\, ,
\end{equation}
with $f=93$ MeV the pion decay constant and $T_{ijl}$, $B^j_l$, $\phi ^k_m$
$SU(3)$ tensors which account for the antidecuplet states, the octet of
$\frac{1}{2}^+$ baryons and the octet of $0^-$ mesons, respectively
\cite{Lee:2004bs}. The second term is given by
\begin{equation}
\label{L2}
{\cal L}_2 = \frac{1}{2 f} \tilde{g}_{\bar{10}}  \epsilon^{ilm} \bar{T}_{ijk}
(\phi \cdot \phi)^j_l B^k_m \, ,
\end{equation}
which couples two mesons in $L=0$ to the antidecuplet and the baryon and they
are in $I=0$ for the case of two pions.  From the Lagrangian terms of Eqs.
(\ref{L1}, \ref{L2}) one can obtain, after some $SU(3)$ algebra,
the transition amplitudes from any member of the
antidecuplet to the different $MMB$ channels to which it couples, in particular
$N^*\to \pi\pi N$. Taking the central values from the PDG \cite{PDG} for the
$N^* (1710) \to N(\pi\pi, p-\textrm{wave},I=1)$ (which we take from the $\rho
N$ fraction of the $N \pi\pi$ decay) and for the $N^* (1710) \to N(\pi\pi,
s-\textrm{wave},I=0)$, the resulting coupling constants are $g_{\bar{10}}=0.72$
and  $\tilde{g}_{\bar{10}}=1.9$.

The $\Theta^+$ selfenergy associated to the diagram of Fig. \ref{Threebody} is
given by 
\begin{eqnarray}
\label{sigmathreebody}
\Sigma^{(V)}(p) =  
 18 \, \Sigma^V (p;K\pi N) + 18 \, \Sigma^V (p;K\eta N) 
\, ,
\nonumber \\
\Sigma^{(S)}(p) =  
 18 \, \Sigma^S (p;K\pi N) + 2 \, \Sigma^S (p;K\eta N) 
\, ,
\end{eqnarray}
with
\begin{eqnarray}
\label{sigmaj}
\Sigma^j(p) &=& - \int \frac{d^4 k}{(2\pi)^4} \int \frac{d^4 q}{(2\pi)^4}
|t^j|^2 \frac{1}{k^2-m_1^2+i\epsilon} \, \frac{1}{q^2-m_2^2+i\epsilon}
\nonumber \\ & &
\frac{M}{E_N(\vec{k}+\vec{q})} \,
\frac{1}{p^0-k^0-q^0-E_N(\vec{k}+\vec{q})+i\epsilon}
\, ,
\end{eqnarray}
where $j \equiv V,S$ and
$m_1$, $m_2$ are the masses of the mesons in the loop ($K \eta$, $K \pi$).
Eq. (\ref{sigmaj}) stands for the $\Theta^+$ selfenergy at rest. In contrast to
the $KN$ decay channel, the dependence on the $\Theta^+$ momentum is not
relevant. The $t^j$ amplitudes in Eq. (\ref{sigmaj}) are given by
\begin{eqnarray}
\label{ts}
|t^S|^2 &=& \left( \frac{\tilde{g}_{\bar{10}}}{2 f} \right) ^2
\,\,\, ,
\nonumber \\
|t^V|^2 &=& \left( \frac{g_{\bar{10}}}{4 f^2} \right) ^2 \frac{1}{2 M}
\lbrace 
\lbrack E_N(\vec{k}+\vec{q}) + M \rbrack 
\lbrack \omega_1(k)-\omega_2(q) \rbrack ^2
\nonumber \\ & &
+ 2 (\vec{k}\,^2-\vec{q}\,^2) \lbrack \omega_1(k)-\omega_2(q) \rbrack
+ \lbrack E_N(\vec{k}+\vec{q}) - M \rbrack (\vec{k}-\vec{q})^2
\rbrace \,\,\,.
\end{eqnarray}
The implementation of the medium effects is done by including the medium
selfenergy of the kaon and modifying the nucleon propagator, as
done in Section \ref{Sigma_korean}. 
On the other hand, the pion being so light requires a more careful treatment
and here we use, as normally done
\cite{Oset:1981ih,Garcia-Recio:1987ik,Herrmann:1993za,Schuck:jn,Urban:1999im,Chiang:1997di,Cabrera:2004kt}, 
the $p-$wave selfenergy from
$ph$ and $\Delta h$ excitation. It is convenient to write the pion propagator
in terms of its Lehman representation \cite{Garcia-Recio:1987ik} and we have 
\begin{eqnarray}
\frac{1}{q^2-m_\pi^2-\Pi_{\pi}(q^0,{\vec q},\rho)}~=~\int_0^ \infty ~d\omega~
2\omega ~\frac{S_{\pi}(\omega,{\vec q},\rho)}
{{q^0}^2-\omega^2+i\epsilon}~,
\end{eqnarray}
where $S_{\pi}(\omega,{\vec q},\rho)$ is the pion spectral function
\begin{eqnarray}
S_{\pi}(\omega,{\vec q},\rho)~=~ -\frac{1}{\pi} \,
\frac{\textrm{Im}~\Pi_{\pi}(\omega,{\vec q},\rho)}{\left|\omega^2-
{\vec q}\,^2-m_\pi^2-\Pi_{\pi}(\omega,{\vec q},\rho)\right|^2}~.
\end{eqnarray}
By performing the energy integrals in Eq. (\ref{sigmaj}) after the medium
effects are incorporated, we get for the $K\pi N$ intermediate channel the following results
\begin{eqnarray}
\Sigma^j(p)~&=&~\int \frac{d^3 k}{(2\pi)^3}\int\frac{d^3 q}{(2\pi)^3}\int_0^ \infty d\omega~
S_{\pi}(\omega,{\vec q},\rho)\left|t^j\right|^2 \nonumber\\
&&\frac{1}{2\tilde\omega(k)}\frac{M}{E_N({\vec k}+{\vec q})}
\frac{1-n({\vec k}+{\vec q})}{p^0-\tilde\omega(k)-\omega-E_N({\vec k}+{\vec q})-V_N+i\epsilon}~~.
\label{pnkint}
\end{eqnarray}
The $K\eta$ channel contributes little to the $\Theta^+$ selfenergy and since the changes of $\eta$ 
in the medium are very small compared to those of the pion, we disregard this channel to account for the medium 
contributions. 

Once the $\Theta^+$ selfenergy at a density $\rho$ is evaluated, the optical potential felt by the
$\Theta^+$ in the medium is obtained by subtracting the free $\Theta^+$ selfenergy, and hence
\begin{eqnarray}
\tilde\Sigma (p)~=~\Sigma(p,\rho)-\Sigma(p,\rho=0)~.
\end{eqnarray}

We should also note that while the $\Theta^+\rightarrow K\pi N$ decay is
forbidden, in the medium the $\pi$ can lead to a $ph$ excitation and this opens
a new decay channel $\Theta^+ N\rightarrow NNK$, which is open down to 1432
MeV, quite below the free $\Theta^+$ mass. We will show that the width from
this channel is also  very small, but should the $\Theta^+$ free width be of the
order of 1 MeV as suggested in  \cite{Gibbs:2004ji,Sibirtsev:2004bg} the new
decay mode would make the width in the medium larger than the free width.

The integral of Eq. (\ref{pnkint}) is regularized by means of a cut-off. We use
a cut-off of around 700-800 MeV by means of which reasonable results for the
vacuum selfenergy of the antidecuplet, of the order of $7-15$ \% of the
antidecuplet masses, are obtained\footnote{In a recent paper where a QCD sum
rule is used to obtain the mass of an $S=1$, $I=0$, $J^P=\frac{1}{2}^+$ state
made from two diquarks and one antiquark, central masses around 1.64 GeV are
obtained \cite{Eidemuller:2004ra} which welcome an extra attractive contribution of $100-150$ MeV from
the $MMB$ components (heptaquark) as we find.}. 
More specifically for $p^0=1540$~MeV and $\Lambda=700$~MeV we find a contribution
to the free $\Theta^+$ selfenergy of $48$~MeV from the scalar Lagrangian and
$40$~MeV from the vector Lagrangian. We find here that the additional attraction of the
$\Theta^+$ in the medium at $\rho=\rho_0$ is of the same order of magnitude as
the binding created by the mechanism considered in the $\Theta^+$ mass. Similar
conclusions were found in the work of \cite{Miller:2004rj} where, although the
formalisms used are quite different, the works share some basic features,
like the small coupling of the $\Theta^+$ to $KN$ and a sizeable coupling to
$K N \pi$, with $K \pi$ with $K^*$ quantum numbers via the vector Lagrangian,
which in \cite{Miller:2004rj} is realized by a large coupling to $K^* N$.
In Ref. \cite{madrid} a thorough study is done of the free selfenergy of
all states of the antidecuplet due to the two meson cloud, using the same
Lagrangians as here as well as other ones which are allowed by $SU(3)$ symmetry
considerations. Several constraints from phenomenology lead to the Lagrangians
which we use here as the leading ones. The vector Lagrangian is also further
modified in \cite{madrid} to account for the decay of the 
$N^*(1710) \to N(\pi\pi,p-\textrm{wave})$ into the actual $N\rho$ channel quoted
in the PDG. This reduces the strength of the vector Lagrangian contribution, and
hence the numbers obtained here for the binding would be reduced by about 20\%
from these corrections.
%%%%%%%

\begin{figure}
\begin{center}
 \includegraphics[width=0.75\textwidth]{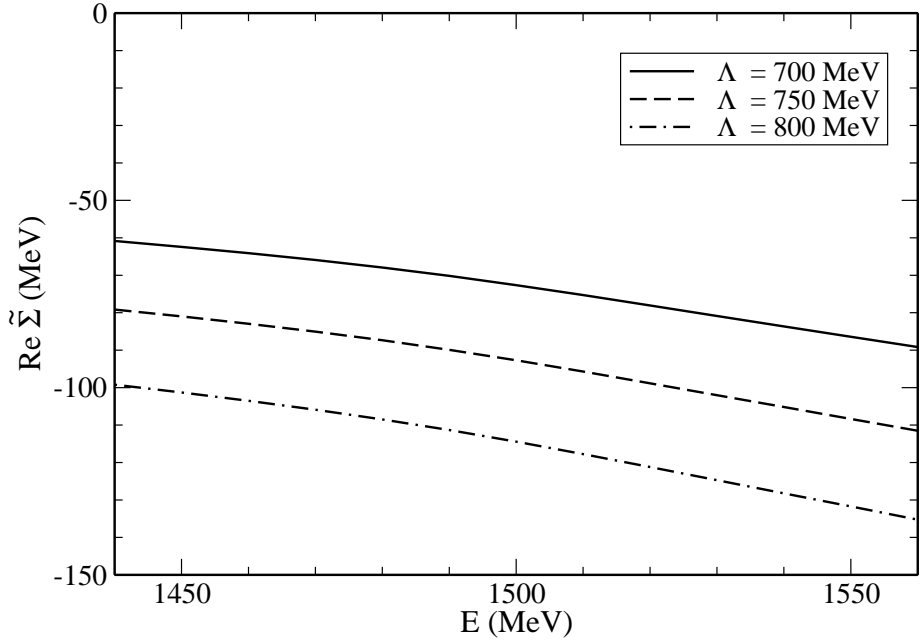}
\caption{Real part of the two-meson contribution to the $\Theta^+$ selfenergy
at $\rho=\rho_0$.} 
\label{Re2meson}
\end{center}
\end{figure}

\begin{figure}
\begin{center}
    \includegraphics[width=0.75\textwidth]{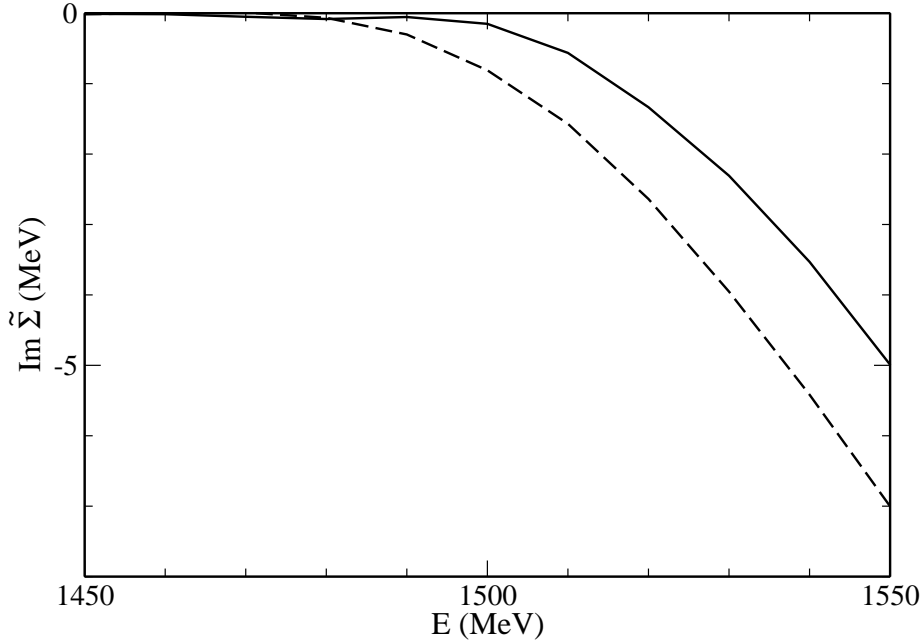}
\caption{Imaginary part of the two-meson contribution to the $\Theta^+$
selfenergy at $\rho=\rho_0$, using the Thomas-Fermi potential (solid line) or
$V_N(r)=-50\rho(r)/\rho_0$~[MeV] (dashed line) for the nucleon potential.}
\label{Im2meson}
\end{center}
\end{figure}

We present the results in Figs. \ref{Re2meson} and \ref{Im2meson}. 
From Fig. \ref{Re2meson} we can see
that the potential for $\rho=\rho_0$ is sizable and attractive and goes
from $-70$ MeV using a cut-off of 700 MeV to $-120$ MeV using 800 MeV. 

Even with the large uncertainties we conclude that there is a sizable
attraction of the order of magnitude of 50-100 MeV at normal nuclear density,
which is more than enough to bind the $\Theta^+$ in any nucleus. In Fig.
\ref{Im2meson} we show the imaginary part of the $\Theta^+$ selfenergy related
to the two-meson decay mechanism for two different nucleon potentials in the
nucleus discussed below. 
We can see that $\Gamma=-2\,\textrm{Im}\,\Sigma$ would be
smaller than 5 MeV for bound states with a binding of $\sim$20 MeV and
negligible for binding energies of $\sim$40 MeV or bigger. This, together with
the small widths associated to the $KN$ decay channel, would lead to $\Theta^+$
widths below 8 MeV, assuming a free width of 15 MeV, and much lower if the
$\Theta^+$ free width is of the order of 1 MeV. In any case, for most
nuclei, this width would be smaller than the separation of the deep levels.

The calculations done here are performed for infinite nuclear matter. In order
to apply the results to finite nuclei we resort to use the local density
approximation, $\rho \to \rho (r)$, with $\rho (r)$ the realistic density
distribution in the nucleus, which is shown to be a good approximation in
\cite{Nieves:1993ev}.

In order to illustrate the point about width and separation of levels we solve the Schr\"odinger equation with two
potentials: $V(r)=-120\rho (r)/\rho_0$  (MeV), and $-60 \rho (r)/\rho_0$ (MeV).
The density $\rho(r)$ is taken from experiment \cite{DeJager:1974dg} for
several nuclei. The results are shown in Table~\ref{tabla}. In a light nucleus
like $^{12}C$ we find several bound states separated by around 20 MeV or more
with both potentials. For medium and heavy nuclei, as in $^{40}Ca$ shown in the
table, we find more bound states and the energy separation is somewhat smaller.

It is  important to remark  that the separation of the deep states is
reasonably bigger than the upper bounds estimated for the width of these states
obtained by considering the $KN$ and the $K\pi N$ decay channels in the medium.
This would make a clear case for the experimental observation of these states.

We have considered two other sources of uncertainty in the calculations. In the
first place we have also included the pion selfenergy due to $2p2h$ excitation
which leads to pion absorption. The pion in the loop in Fig. \ref{Threebody}
cannot be put on shell simultaneously with the $K$ and the nucleon. The pion can
excite a $ph$ and this gives the $K^+ N ph$ decay mode of the $\Theta^+$ which
we have studied. The pion can also lead to $2p2h$ excitation and this would give
a new decay mode, $K^+ N 2p2h$. Since this is a ${\cal O}(\rho^2)$ correction
compared to the ${\cal O}(\rho)$ contribution of the $ph$ decay channel, we can
think a priori that the $2p2h$ channel will be less relevant than the $ph$ one.
The situation is reminiscent of the one nucleon induced and two nucleon induced
$\Lambda$ decay in nuclei \cite{Alberico:1990wg,Ramos:1994xy} where the $\Lambda \to N\pi$ decay is
also forbidden by Pauli blocking. There one finds \cite{Ramos:1994xy} that the two nucleon
induced decay represents a fraction smaller that 20\% of the $ph$ one. In the
present case there is even less energy left for the pion as an average than in
$\Lambda$ decay, and Pauli blocking is also more effective, thus we should
expect smaller results. This is indeed the case as we find in actual
calculations. For this purpose we include the selfenergy given in \cite{Ramos:1994xy},
\begin{equation}
\label{2p2hselfenergy}
\Pi^{2p2h}_{\pi} = -4\pi \vec{q}\,^2 {\cal C}_0^* \rho^2 \,\,\, ,
\end{equation}
with ${\cal C}_0^* = (0.105 + i 0.096) m_{\pi}^{-6}$, which is obtained from
pionic atoms and is modified in \cite{Ramos:1994xy} to account for the different phase
space offered by the off-shell pions which we find in the present case. The
result of the calculation is that the strength of the real part of the $\Theta^+$ selfenergy in
the medium decreases by a few MeV and the imaginary part increases less than 5\%.
Hence, the effect of including this new decay channel is negligible considering
the large uncertainty from other sources.

The other element considered has to do with the nucleon binding. In Eq.
(\ref{SigmaAll}) we took the Thomas-Fermi potential for the nucleon. Now we take
a standard potential,
\begin{equation}
\label{otherVN}
V_N = -V_0 \rho/\rho_0\,\,\,;\\\ V_0=50~\textrm{MeV} \,\,\,,
\end{equation}
and recalculate the results. The real part increases by a few MeV and the
imaginary part, plotted in Fig. \ref{Im2meson}, shows some changes with
respect to that obtained with the Thomas-Fermi potential, still leading to a
width of only 5~MeV for 20~MeV binding.

\begin{table}
\begin{center}
\label{tabla}
\begin{tabular}{|c|c||c|c|}
\hline 
\multicolumn{2}{|c||}{$V=-60$ MeV $\rho/\rho_0$}&
\multicolumn{2}{|c|}{$V=-120$ MeV $\rho/\rho_0$}\\
\hline
$E_{i}$ (MeV), $^{12}C$ &
$E_{i}$ (MeV), $^{40}Ca$ &
$E_{i}$ (MeV), $^{12}C$&
$E_{i}$ (MeV), $^{40}Ca$ \\
\hline
\hline 
-34.0 (1s)&
-42.6 (1s)&
-87.3 (1s)&
-98.2 (1s)\\
%\hline 
-14.6 (1p)&
-30.9 (1p)&
-59.5 (1p)&
-83.3 (1p)\\
%\hline
-0.3 (2s) &
-18.7 (1d)&
-32.0 (2s)&
-67.5 (1d)\\
%\hline
           &
-17.9 (2s) &
-31.9 (1d) &
-65.9 (2s)\\
%\hline
&
-6.3 (1f)&
-8.6 (2p)&
-50.8 (1f)\\
&
-5.6 (2p)&
-5.6 (1f)&
-48.5 (2p)\\
%\hline
&
&
&
-33.5 (1g)\\
&
&
&
-31.1 (2d)\\
&
&
&
-30.4 (3s)\\
&
&
&
-15.9 (1h)\\
&
&
&
-14.2 (2f)\\
&
&
&
-13.8 (3p)\\
&
&
&
-0.5 (4s)\\
\hline
\end{tabular}
\caption{Binding energies of $\Theta^+$ in $^{12}C$ and $^{40}Ca$.}
\end{center}
\end{table}

\section{Conclusions}
We have evaluated the selfenergy of the $\Theta^+$ in the nuclear medium
associated to the $KN$ decay channels and the $MMB$ decay channels of the
$\Theta^+$ partners in the antidecuplet. We obtain a small potential
associated to the $KN$ decay, even assuming a large free width of around 15 MeV
for the $\Theta^+$, but at the same time we also show that Pauli blocking and
the decreased phase space from the $\Theta^+$ binding decrease appreciably the
$\Theta^+$ width in the nucleus from the $KN$ decay.

On the other hand, we find a large attractive $\Theta^+$ potential in the nucleus associated to the two meson
cloud of the antidecuplet. A new decay channel opens for the $\Theta^+$ in the medium, 
$\Theta^+ N\rightarrow NNK$, but the width from this new channel, together with the one from $KN$ decay, is
still small compared to the separation of the bound levels of the $\Theta^+$ in light and intermediate nuclei
(very large nuclei like $^{208}Pb$ would have the states too packed to prove efficient in the detection of these
states). 

In reaching the former conclusions there are several assumptions done.
\begin{enumerate}
\item The $\Theta^+$ is assumed to be $1/2^+$ associated to an $SU(3)$ antidecuplet;
\item The $N^*(1710)$ is supposed to couple largely to this antidecuplet;
\item The Lagrangians have been chosen to 
reproduce the $N(\pi\pi,p-\textrm{wave}, I=1)$ and $N(\pi\pi,
s-\textrm{wave}, I=0)$ decay mode of $N^* (1710)$ by imposing $SU(3)$ symmetry
with a minimal number of derivatives in the fields.
\item Some values of the cut off have been chosen to obtain reasonable numbers for the free $\Theta^+$
selfenergy;
\item The average value of the $N^*(1710)$ width and the partial decay ratios, which
experimentally have large uncertainties, have been taken to fix the
couplings of the antidecuplet to the baryon octet and the two meson octets.
\end{enumerate} 

It is clear that with all these assumptions one must accept a large uncertainty
in the results. So we can not be precise on the binding energies of the
$\Theta^+$. However, the order of magnitude obtained for the potential is such
that even with a wide margin of uncertainty, the conclusion that there would be
bound states is quite safe. In fact, with potentials with a strength of 20 MeV
or less one would already get bound states. Furthermore, since the strength of
the real part and the imaginary part from the $NKph$ decay are driven by the
same coupling, a reduction on the strength of the potential would also lead to
reduced widths such that the principle that the widths are reasonably smaller
than the separation between levels would be saved.

The work done here provides thus a sensible case in favor of the existence of
bound $\Theta^+$ states in nuclei  which should spur experimental work in this
area.  

\section*{Acknowledgments}
We would like to thank T.~Hyodo for useful comments. 
This work is partly supported by DGICYT contract number BFM2003-00856,
and the E.U. EURIDICE network contract no. HPRN-CT-2002-00311. D.~C.
acknowledges financial support from MCYT and Q.~B.~Li acknowledges support from
the Ministerio de Educaci\'on y Ciencia in the program of Doctores y
Tecn\'ologos Extranjeros.

\vfill
\end{document}